\documentclass[runningheads,a4paper]{llncs}

\usepackage{amssymb}
\usepackage{graphicx}
\usepackage{epstopdf}
\usepackage{xcolor}
\usepackage{subfigure}
\usepackage{algorithm}
\usepackage{algorithmic}
\usepackage{url}
\urldef{\mailsa}\path|{zcqin,|
\urldef{\mailsc}\path|taowan}@buaa.edu.cn|
\newcommand{\keywords}[1]{\par\addvspace\baselineskip
\noindent\keywordname\enspace\ignorespaces#1}

\begin{document}

\mainmatter  % start of an individual contribution

\title{Motif Iteration Model for Network Representation}

\titlerunning{Motif Iteration Model for Network Representation}

\author{Lintao Lv$^1$%
%\thanks{.}%
\and Zengchang Qin$^1$ \and Tao Wan$^2$
%\and Frank Holzwarth\and\\
%Anna Kramer\and Leonie Kunz\and Christine Rei\ss\and\\
%Nicole Sator\and Erika Siebert-Cole\and Peter Stra\ss er
}
\authorrunning{Lv, Qin and Wan}

\institute{$^1$Intelligent Computing and Machine Learning Lab\\
School of Automation Science and Electrical Engineering \\
Beihang University, Beijing, 100191, Beijing, China \\
$^2$School of Biological Science and Medical Engineering \\
Beihang University, Beijing, 100191, Beijing, China \\
\mailsa
\mailsc\\
%\url{http://www.springer.com/lncs}
}


\toctitle{Lecture Notes in Computer Science}
\tocauthor{Authors' Instructions}
\maketitle

\begin{abstract}
  Social media mining has become one of the most popular research areas in Big Data with the explosion of
  social networking information from Facebook, Twitter, LinkedIn, Weibo and so on.
  Understanding and representing the structure of a social network is a key in social media mining.
  %Classically, a network is repreente
  In this paper, we propose the {\em Motif Iteration Model} (MIM) to represent the structure of a social network.
  As the name suggested, the new model is based on iteration of basic network motifs.
  %Based on the idea of the most primitive network motifs, we propose the MIM to describe the complex network.
  %Further on this model,
  In order to better show the properties of the model, a heuristic and greedy algorithm called {\em Vertex Reordering and Arranging} (VRA) is proposed by studying the adjacency matrix of the three-vertex undirected network motifs.
  The algorithm is for mapping from the adjacency matrix of a network to a binary image, it shows
   a new perspective of network structure visualization.
  %A complex network can be linearly reordered and arranged and sequentially with vertex symmetry. That makes the images of adjacency matrix have obvious %patterns.
  In summary, this model provides a useful approach towards building link between images and networks and offers a new way of representing the structure of a social network.
\end{abstract}

\section{Introduction}
Over the past a few years, there has been an explosion of interests in social media (network) mining with the increase popularity of
online social networking services like Facebook, Twitter, Weibo and so on.
In the study of social networks, the most fundamental idea in social network research is that a node's position in a network determines in part the opportunities and constraints that it encounter~\cite{int0}.
From transcriptional regulation networks, computer networks to electrical circuits networks, network motifs is regarded as
recurring circuits of interactions from which the networks are built.
Network motifs, depend on a small set of recurring regulation patterns, inspire more research on the networks~\cite{int1,int2,int3,int4}.

%They have recently gathered much attention as a useful concept to represent structure of complex networks. However, it should be noted that the detection of network motifs is still a challenge thing.
Adjacency matrix is a square matrix representing the graph of a finite network. The binary value in adjacency matrix indicates whether the vertices are connected or not.
On the other hand, the largest eigenvalue of the network adjacency matrix has emerged as a key value for study of a variety of dynamical networks.
This allows the degree of a vertex to be easily found by taking the sum of the values in either its respective row or column in the adjacency matrix. Manipulating the adjacency matrix is the most direct way to study the complex network.
In literatures, there are two major kinds of models for learning network structures: (1) count-based motifs methods, such as Generative Model Selection for Complex Networks (GMSCN) \cite{int5} and (2) Global adjacency matrix methods, such as Structural Perturbation Method (SPM) for link predictability of complex networks~\cite{int6}. Currently, both models suffer significant drawbacks. While methods like GMSCN efficiently leverage statistical information, they do relatively poorly on the globe structure. SPM may do better on the analogy task of globe structure, but they poorly utilize the statistics of the graphlets.

In this work, we analyze the both model properties to produce linear mapping between the social network structure and the adjacency matrix binary ({\em amb}) image .
We also introduce the Motifs Iteration Model (MIM) for globe network structure and use an algorithm to indicate the relevance of the network structure and the {\em amb} image.
%The MIM is structured with the iteration of basic network motifs.
Under certain constraints, it can be a good representation of the real network structure, including the local motifs unit information and the overall frame information.
After discussing the relationship between the adjacency matrix and the network motifs, we present a heuristic algorithm VRA.
This method consists of two important parts, including reordering of the complex network vertex and replacing the sorted results symmetrically.
%The rest of this paper is organized as follows. Section II reviews the related work. Section III presents new topology view of networks. Section IV is dedicated for NRA. Section V describes the simple study about network models and real networks. VI concludes the paper.

\section{Related Work}

%\subsection{Network generation models}
Considering the structure,
there are mainly four types of well-studied networks:
random networks (Erd\"{o}s-R\'{e}nyi  model)~\cite{relER},
nearest-neighbor coupled networks (NCN model)~\cite{relNCN},
small world networks (Watts-Strogatz model)~\cite{relWS}
and scale-free networks (Barab\'{a}si-Albert model)~\cite{relBA}.

%\begin{itemize}
(1) Given a { random network} with $N$ vertices, there can be $C^N_2$ edges, and the network from which we randomly connect $M$ edges is called random network.
    The links between the vertices in the network are random, the characteristics of the network is also random.
 (2) For the nearest-neighbor coupled networks, each vertex in the network is linked to a fixed number of vertices around the vertex.
    NCN network has a stable degree distribution.
 (3){ The small-world networks} is another classic network. It has a small path length and high clustering properties, started with a regular network, such as NCN, and we can obtain a small-world network by re-routing some edges randomly.
 (4) A typical feature of the scale-free network is that most vertices in a network are connected to only a few vertices, and very few vertices are connected to a very large number of vertices.
    The existence of such a critical vertex makes the scale-free network a strong ability to withstand unexpected failures, but in the face of collaborative attack is vulnerable.

%\end{itemize}

%We have only selected four common complex network models for our study.
%Other generative models are also used in related model selection methods, but we have not utilized.

%\subsection{Related applications}

In the field of complex networks,
Janssen~\cite{rel1} tries to classify the multiple network models by using a broad array of features, include the frequency counts of small subgraphs as well as features capturing the degree distribution and small world property, in order to make a better judgment on the real network.
In the study of compressing the large-scale network, Liakos~\cite{rel2} improves the state-of-the-art method for graph compression by exploiting the locality of reference observed in social network graphs.
They apply the Layered Label Propagation(LLP)~\cite{relLLP} algorithm on the origin social network graphs.
More complex a network is, more accurate results will be.
But they can not visualize the good performance of the model, only from theoretical and experimental verification of the superiority of the method.
Although the network motifs and adjacency matrix in the application of the social networks were achieved good results, the use of combination of the two studies to social networks is relatively unexplored.

\section{Topology Of Networks}

\subsection{Motifs Iteration Model}

Most study about complex networks mainly analyze their global statistical characteristics, such as small world features, scale-free features.
However, in addition to these global features, the characteristics of the basic elements of each type of network is also very important.
The network motifs are the key patterns of interaction in complex networks, which are more common in complex networks than in random networks~\cite{motifs}.
Network motifs, as one of the best conditions of complex network, reveals the basic information of structure or basic building blocks of most complex networks in the real-world.

In different types of network structures, the number and type of different motifs vary widely.
In this paper, to simplify the problem, we mainly study the network structure of undirected graphs.
We will select the graphlets of the three-node, four-node undirected graphs in our study.
Since the network element can only express local information in the network structure, we want to amplify the representation of the network motifs.
The most basic element in the network motifs is a vertex, and we replace it with another network motifs.
As each member in the family will eventually form their own family, this alternative approach is a typical iterative approach.
The replaced network motifs is defined as a second-level network motifs.
We then will obtain a higher level of network motifs.
This is referred to as Motifs Iteration Model (MIM).

\begin{figure}
\centering
\includegraphics[width=3.4in]{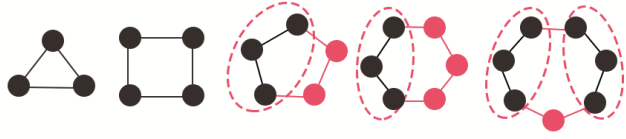}
\caption{Displays the Motifs Iteration Model of the networks under different number of nodes. There is an NCN network with parameter $k = 2$. Black nodes and segments are denoted as first-level motifs, red nodes and line segments represent second-layer motifs. The link between the second-layer motifs is called the second-link.}
\end{figure}

We only need to ensure that each layer of the network motifs are consisted of node, three-node motifs and four-node motifs.
The details of the MIM are shown in Figure 1.
From left to right are the network graph with different node number, and each node in the network connected with the adjacent node.
Black node as a first-level network motifs, red node with red dashed ellipse as second-layer network motifs.
The black and red connections represent the element links of the first-layer and second-layer network motifs, respectively.
Finally, we can use MIM of network to form any complex network structure easily.
Prove as follows:

\begin{proof}
Let $n$ be the size of the networks.
We omit the situation of $ n < 6 $, because this is clearly held.
And define vertices as $V$, and the basic network motifs as $M(n) = (x, y, z)$ or $M(n) = (x, y, z, p)$. Where $ x, y, z, p$ can be a single vertex or motifs.
So $M(7) = (M(3), M(3), 1) = (M(4), 1, 1, 1)$.

We assume that the number of vertices in a complex network is $k$, we can use the above formula.
Which is:
\begin{equation}
   M(k) = (\alpha, \beta, \gamma) | (\alpha, \beta, \gamma, \delta)
\end{equation}

So we can introduce the network vertex count for the case of $k + 1$.
Here, we will discuss two cases. If $ M(k) = (\alpha, \beta, \gamma)$, so $ M(k + 1) = (\alpha, \beta, \gamma, 1) $.
If $ M(k) = (\alpha, \beta, \gamma, \delta)$, $ M(k + 1)$ will become $ ((\alpha, \beta, \gamma), \delta, 1) $.
It is worth noting that: $\alpha, \beta, \gamma$ stand for the elements randomly.
Based on this derivation, it is proved that all networks will be represented by the MIM.
\qed
\end{proof}

\subsection{Standard Motifs Iteration Model}

In the process of building the model based on MIM, we will still face the choice, including the motifs level and node placement.
For medium-sized networks, there will be tens of thousands of cases.
In order to better represent the network structure, we define Standard Motifs Iteration Model (SMIM) that reducing the choice of building a network structure using MIM.
The biggest difference from the previous model is that the SMIM adds additional constraints.
\begin{itemize}
\item The nodes in adjacent locations are preferentially linked.
\item The network motifs at the same level are preferentially linked.
\end{itemize}
The first constraint is to ensure that the elements of the motifs in each network are close, so that the local information of the network structure is better preserved.
The other constraint is for having good uniformity in the adjacency matrix binary{\em amb} image.
Based on this model, we need to re-evaluate the capabilities of SMIM for the network.
We can prove this as following:

\begin{proof}
%We omitted some of the derivation, because they have already appeared before.
When $ M(k) = (\alpha, \beta, \gamma)$ is true, $ M(k + 1) = (\alpha, \beta, \gamma, 1)$ is true at the same time.
When $ M(k) = (\alpha, \beta, \gamma, \delta)$ is true, since $k$ is greater than 7, there must be an element in $ \alpha, \beta, \gamma, \delta $  that is not a single vertex.
\begin{itemize}
         \item If $ M(k) = (\alpha, 1, 1, 1)$, so $ M(k + 1) = (\alpha, M(3), 1) $;
         \item If $ M(k) = (\alpha, \beta, 1, 1)$, so $ M(k + 1) = (\alpha, \beta, M(3)) $;
         \item If $ M(k) = (\alpha, \beta, \gamma, 1)$, so $ M(k + 1) = ((\alpha, \beta, \gamma), 1, 1) $;
         \item If $ M(k) = (\alpha, \beta, \gamma, \delta)$, so $ M(k + 1) = ((\alpha, \beta, \gamma), \delta, 1) $;
\end{itemize}
\qed
\end{proof}
Through the study of the proof, we can find: the most advanced motifs in the complex network is showing the three nodes, the four nodes change alternately.
And in the later work, the symmetry of this property in the network motifs and network adjacency matrix combination process will play an important role.
In order to specify the model, the following describes the network model with a vertex number of 13:
\begin{equation}
   M(13) = ((1, 1, 1), ((1, 1, 1), (1, 1, 1), (1, 1, 1)), 1)
\end{equation}

\subsection{Adjacency Matrix Of Motifs}
The structure of the undirected network decides adjacency matrix symmetry.
Each element in the adjacency matrix corresponds to a single pixel, the adjacency matrix can correspond to an $n \times n$  binary image.
In the adjacency matrix, if there is a connection between the two vertices, the corresponding element is 1, otherwise 0.
Element $1$ corresponds to white plot, and element $0$ is a black plot.
Therefore, based on the adjacency matrix, we can plot an {\em adjacency matrix binary (amb)} image.
For a fixed network motif, we hope to find the unique representation of the {\em amb} image.
If we can solve the mapping problem of the undirected motifs and the adjacency matrix binary image, we can observe the complex network from the network adjacency matrix binary image of the real network.

\begin{figure}
\centering
\includegraphics[width=3.2in]{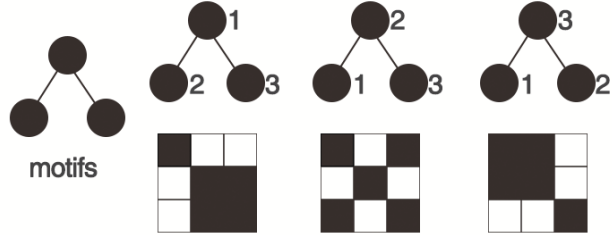}
\caption{This is an example of a three-node motifs with its corresponding {\em amb} image. The first line is the same motifs structure with different node indices. The second line is network's {\em amb} image. Though the structure of the network is identical, but the {\em amb} images are different, we need one unique image for the same network structure regardless node indices. }
\label{fig:ambdif}
\end{figure}

In the SMIM, three network motifs are used as essential components.
In other words, the state of more than four motifs is only regarded as generated structure from basic three-vertices.
However, even with a simple three-vertices network of the same structure, different vertex indexing will result in different {\em amb} images. For example, see Fig. \ref{fig:ambdif}.
%Three vertices of the network motifs corresponding to the adjacency binary matrix image shown in Figure 3.

In Fig. 2, the left side shows the network motifs we used, in order to illustrate the importance of the vertex number, we
did not use fully connected three-node network motifs.
Because when the network with fully connected, the vertex indexing will not have any affect on the {\em amb} image.
%The three markings on the top right are selected from all the markings because the corresponding {\em amb} image is different.
There are three cases are named (a), (b) and (c), from the left to right.
All the cases have the same structure,
but for their {\em amb} images, they are not symmetrical.
The {\em amb} image of case (b) is symmetric and symmetrical with the standard multi-layer network motifs.
Based on the advantages of motifs, we will examine the correspondence between the network motifs and the {\em amb} image of case (b).

\subsection{Vertex Reordering And Arranging}

Inspired by the motifs of the three-node example, we can arrange the important vertices in the middle, and the secondary vertex is in the adjacent position.
In accordance with this idea, propose the following Vertex Reordering and Arranging (VRA) algorithm.
\begin{algorithm}
  \caption{Vertex reordering and arranging}
  \label{alg1}
  \begin{algorithmic}[1]
  \REQUIRE Adjacency matrix $M$ of the given network.
  \ENSURE New adjacency matrix $\hat M$.
  \STATE Calculate the degree $d_i$ of $i$th vertex.
  \STATE Compute degree list $\hat {d_i}$.
  \STATE Define empty list $X$ and $max = max(\hat {d_i}) - 1$.
  \FORALL {$ \hat {d_i}  = max$ }
            \STATE all possible values are denoted as $\Omega$.
            \STATE Select $a$ in $\Omega$ randomly.
            \FORALL { len$(a) - 1$ }
                    \STATE Append $a$ in $X$ list.
                    \IF{$b$ in $\Omega$ but not in $X$ and $b$ satisfies the priority }
                           \STATE $a = b$ and break.
                    \ENDIF
                    \STATE Select $a$ in $(\Omega - X)$ randomly and append $a$ in $X$.
            \ENDFOR
            \STATE $max = max - 1$
  \ENDFOR
  \STATE The odd-numbered elements in $X$ are placed in front of the list in turn.
  \STATE $X \times X$ constitute the $\hat M$.
  \FORALL { $(i,j)$ in vertexes }
            \STATE $\hat M[i][j] = M[X[i]][X[j]] $
  \ENDFOR
  \end{algorithmic}
\end{algorithm}

In the above algorithm, more choices are provided to satisfy the priority order.
The implementation of the code in this article is based on the interconnection of nodes to be ordered firstly.
VRA is an algorithm based on sorting and special placement, and it is a heuristic algorithm essentially.

\begin{figure}
\centering
\subfigure[]{
\begin{minipage}[t]{0.45\linewidth}
\centering
\includegraphics[width=2in]{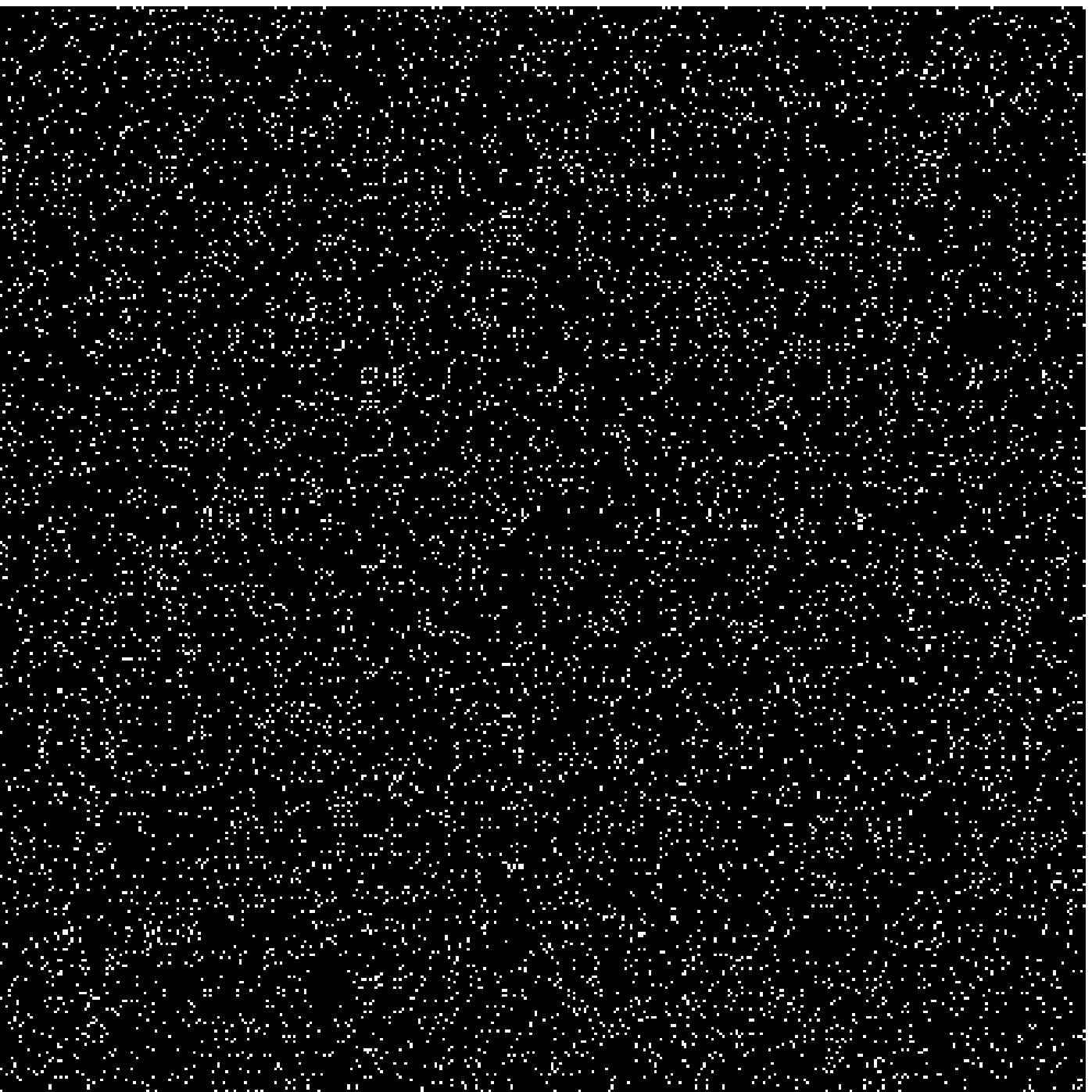}
\label{fig:side:a}
\end{minipage}
}
\subfigure[]{
\begin{minipage}[t]{0.45\linewidth}
\centering
\includegraphics[width=2in]{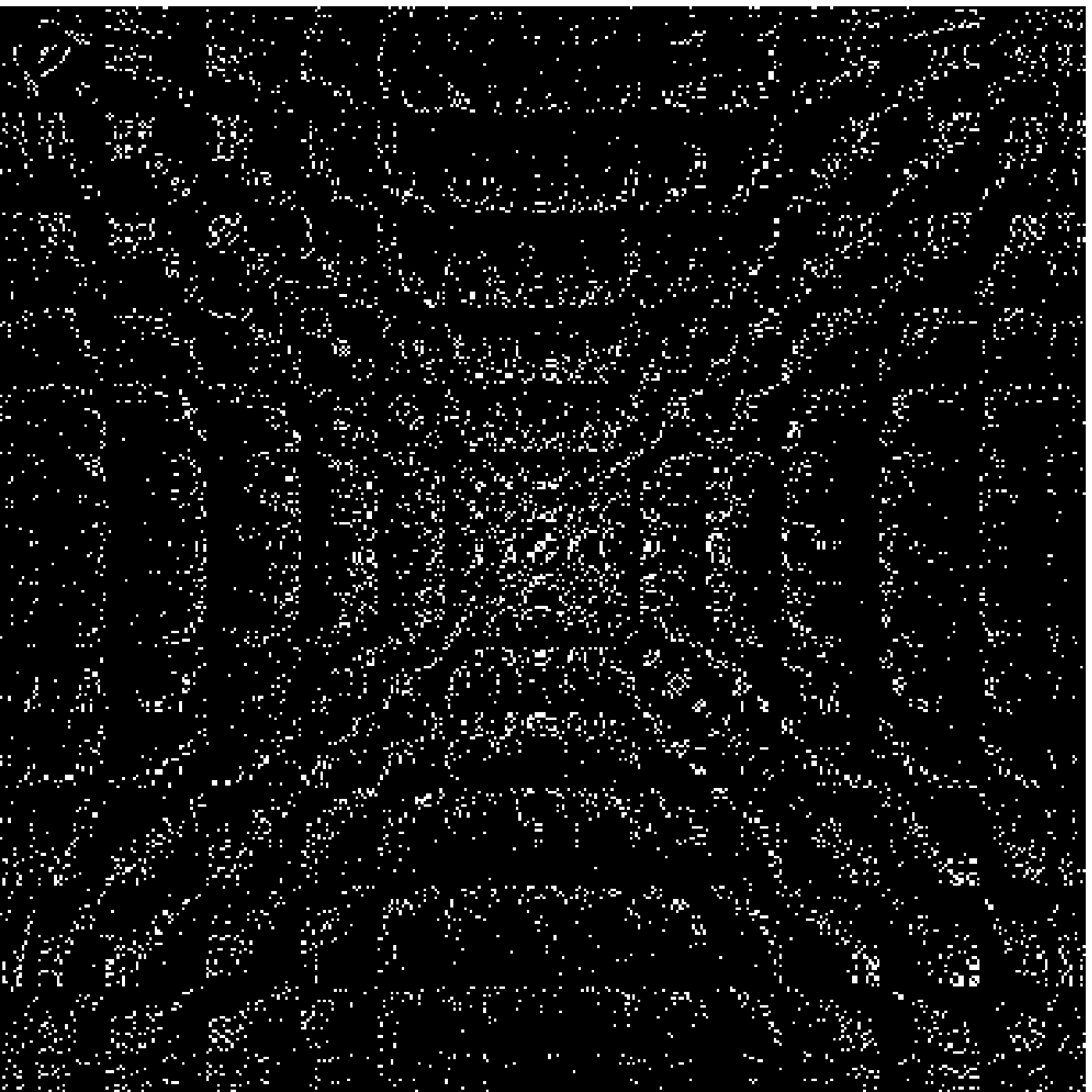}
\label{fig:side:b}
\end{minipage}
}
\caption{An illustration of the VRA effect for small-world networks. (a) and (b) represent the {\em amb} image after random generation and VRA processing respectively.}
\label{fig:vra}
\end{figure}

In Fig. \ref{fig:vra}, it shows that the algorithm has good performance in identifying the network structure.
With a rough look, two images are completely different.
However, two {\em amb} images represent the same network structure.
The example shown in Fig. \ref{fig:vra} is a WS network with parameters (200, 20, 0.1).
Representing 200 vertices in the network, each of which is connected to 20 vertices adjacently to the surroundings, and the presence of 0.1 probability fluctuations in the connection.
The image on the left-hand side is the {\em amb} image corresponding to the number of the vertex of the network, and the image on the right-hand side is the {\em amb} image after the VRA processing.

It is easy to see that the right-hand side image has a clear pattern.
%More specific information instructions: in the right-hand side image, the main diagonal and vice diagonal highlights.
This is because in the SMIM, the adjacent three vertices are small groups, and the small group of elements in the algorithm is adjacent, so the main diagonal is regular.
On the other hand, left and right symmetry of small groups are closely related, so the same data on the diagonal also have the law.
And the white point that is almost parallel to the image is the display form of the relationship between the small group and the large group.
It can also be said that VRA method can better discover the nature of the network.
Now we have established a network and the corresponding relationship between the image.
Combined with SMIM and VRA, we will be able to apply the method of image processing to the application of social networks.

\section{Experimental Studies}

%\subsection{Network identification}
In order to test the validity of our new model to capture the properties of complex networks, we classify the {\em amb} image into different network classes based on the convolutional neural network (CNN)~\cite{CNN} method.
We select the ER network, NCN network, WS network and BA network as our algorithm in the simulation of network classification of test data.
The network is represented by an {\em amb} image after applying the VRA algorithm, and the network type is used as the label of the image as the input of the CNN.
By training the neural network model, we try to find a good network type classifier.

In the data set, different parameters are set for each network.
The number of vertices in the network here is fixed with 100.
In order to test the rendering ability of for complex network's {\em amb} image, we selected the experiments without VRA as a baseline methods.
In order to reflect the network type classification with and without the VRA method, precision and recall are used for measuring the performance. The results are shown in Table 1 and 2, respectively.

\begin{table}
\caption{Precision and Recall of the classifier without VRA.}
\begin{center}
\begin{tabular}{ccccc}
\hline
\hline
\   & \textbf{ER}  & \textbf{NCN} & \textbf{WS} & \textbf{BA} \\
\hline
Precision & 97.5\% & 84.3\% & 98.7\% & 100\%\\
Recall    & 99.0\% & 100\% & 79.0\% & 100\%\\
\hline
\end{tabular}
\end{center}
\end{table}
\begin{table}[htbp]
\caption{Precision and Recall of the classifier with VRA.}
\begin{center}
\begin{tabular}{ccccc}
\hline
\hline
\   & \textbf{ER}  & \textbf{NCN} & \textbf{WS} & \textbf{BA} \\
\hline
Precision & 100\% & 100\% & 100\% & 100\%\\
Recall    & 100\% & 100\% & 100\% & 100\%\\
\hline
\end{tabular}
\end{center}
\end{table}

The VRA processed network classifier shows a strong classification capability.
In the processing of the model training, 100\% accuracy can be achieved.
The difference between two groups of experiments is that the resolution ability of random networks and some small-world networks.
With the change of probability, the small world network's {\em amb} image is very similar with that of the random network.
However, network with VRA can be very easy to avoid this problem.

In the real large-scale social network category identification processing, we found that a social network is not simply generated by one standard network.
In order to better describe the reality of the network, we propose a rough solution.
Based on standards community detection~\cite{Dectection}, each community need to identify what kind of network it is.
Finally, the network can be described as a combination of multiple basic models.
\begin{equation}
    \textbf{D(x)} = 0.5  \textbf{T}(\textbf{x}) + 0.5  \sum_{i=1}^K{\lambda_{i}  \textbf{T}(\textbf{x}_i)}
\end{equation}

Where $\textbf x$ is the entire network, $\textbf x_i$ represents each small community network.
The number of communities found after the standard network community is $k$, and $ \lambda_{i} $ represents the weight.
The $\textbf D()$ function and the  $\textbf T()$ function denote the functions describing the network and the network type function, respectively.

For example, the Zachary¡¯s Karate Club is a social network of friendships between 34 members of a karate club at a US university in the 1970s~\cite{dataZac}.
Need to pay attention to the process of dealing with the real network, the use of the classifier is through the number of vertex 50 training from the network.
When dealing with a network with fewer than 50 network points, we need to put the {\em amb} image corresponding to the network to be tested in the center and the rest to use the black supplement.
The decomposition of this network is as follows: scale-free networks and small-world networks, and a small number of nearest-neighbor coupled networks.
\begin{equation}
    \textbf{D(Zachary)} =  0.50 \textbf{BA} + 0.43 \textbf{WS} + 0.07 \textbf{NCN}
\end{equation}

\section{Conclusions}

Although either network motif and adjacency matrix is well used in studies of social networks and had achieved good results,
how to use both of them is relatively unexplored.
Through the study of network structure, by using motifs, we propose MIM and SMIM models.
Through theoretical exposition and experimental verification, the networks using VRA can present best patterns of adjacency matrix images.
Based on an image representation of the network structure, some image processing or pattern recognition (e.g. convolution neural networks) can be used to study network structures. This leaves a lot of possibilities for our future research.
%Through the combination of images and social networks, we will have a new breakthrough in social networking research.

\subsubsection*{Acknowledgments.} This work is supported by the National Science Foundation of China Nos. 61401012 and 61305047.

\end{document}